\documentclass[allclo]{FBSart}
\usepackage{epsfig,psfig}

\newcommand{\gL}{\Lambda}
\newcommand{\be}{\begin{equation}}
\newcommand{\ee}{\end{equation}}
\title{Dissociation of hadrons in quark matter within finite temperature
         field theory approach on the light front\footnote{Presented by S. Mattiello at Light-Cone 2004, Amsterdam, 16 - 20 August}} 
\author{S. Mattiello and M. Beyer}
\institute{Fachbereich Physik, Universit\"at Rostock, D-18051
  Rostock, Germany\\ e-mail: {\tt stefano.mattiello@uni-rostock.de}}
\runningauthor{S. Mattiello}
\runningtitle{LC 2004}
\begin{document}
\maketitle
\begin{abstract}
We present a relativistic three-body equation to investigate the properties of nucleons in hot and dense nuclear/quark matter. Within the light front approach we utilize a zero-range interaction to study the three-body dynamics. The relativistic in-medium equation is derived within a systematic Dyson equation approach that includes the dominant medium effects due to Pauli blocking and self energy corrections. We present the in-medium nucleon mass and calculate the dissociation of the three-body system.
\end{abstract}
\section{Introduction}

Exploring the phase diagram of the quantum chromodynamics is a theoretically challenging task.
Lattice calculations at zero-density give evidence of chiral and confined-deconfined phase transitions at a temperature $T_c$ of about 150 to 175 MeV~\cite{Karsch:2000vy}.
Experimentally the phase diagram is accessible by heavy ion collisions and is related to astronomical observations of, e.g., neutron stars. 
New methods to extend lattice calculations at small chemical potential are developed, i.e. multiparameter reweighting, Taylor expansion at $\mu\simeq 0$ and imaginary chemical potential, but the validity of these methods is limited to a region $\mu\lesssim T$~\cite{Katz:NPPS129}. For larger densities effective approaches indicate an extremely rich phase diagram, i.e. showing a quark-gluon plasma, a color superconductivity and a color-flavor locking phase~\cite{Alford:2001dt}.
The light front quantization makes it possible to investigate quantum field theory in a Hamiltonian formulation~\cite{Dirac:RMP21}.
This allows us an application to systems at finite temperature and density~\cite{Mattiello:chiral}.
In the vicinity of the chiral phase transition three-quark correlations should play an important role~\cite{Beyer:2001bc}.
The finite temperature field theory on the light cone allows us to investigate the three-quark correlations in a systematic way within the Dyson equation approach.
The relativistic three-body problem has been investigated utilizing the light front form with a contact interaction, that provides a simple, but important limiting case for short-range forces~\cite{Frederico:1992uw,Carbonell:2002qs,Mattiello:TN03}. 

In this paper we present first the isolated relativistic three-particle system using an invariant cut-off $\Lambda$.
After that we consider the properties of the proton in hot and dense quark matter and calculate the dissociation for different values of the cut-off parameter.

\section{Isolated case}
For the time being we consider bose-type relativistic equations.
The two-body propagator $t(M_2)$, given by~\cite{Frederico:1992uw}
\begin{equation}
t(M_2)=\left(i\lambda^{-1} - B(M_2)\right)^{-1},
\label{eqn:tau}
\end{equation}
is the input for the three-body calculation. $B(M_2)$ corresponds to a loop diagram, that in the rest system of the two-body system is given by
\begin{equation}
B(M_2)=-\frac{i}{2(2\pi)^3} \int \frac{dx d^2k_\perp}{x(1-x)}
\frac{1}{M_2^2-M_{20}^2},
\label{eqn:B}
\end{equation}
where $M_{20}^2=(\vec k_\perp^{~2}+m^2)/x(1-x)$ and $x=k^+/P^+_2$. The integral shows a logarithmic divergence. In
order to investigate the three-body bound state equation even if no two-body bound state exists, we invoke an invariant cut-off $\gL$~\cite{Mattiello:TN03}. The requirement is that the mass of the virtual two-body subsystem is smaller than the cut-off, i.e. $M_{20}^2<\gL^2$ and hence $t\rightarrow t_\gL$. In the three-body equations we introduce a similar regularization for the mass
of the virtual three-particle state. That leads to a parametric dependence of the vertex function $\Gamma$ on the cut-off $\gL$,
\begin{eqnarray}
\Gamma_\Lambda(y,\vec q_\perp) &= &\frac{i}{(2\pi)^3}\ t_\Lambda(M_2)
\int_{0}^{1-y} \frac{dx}{x(1-y-x)}\nonumber\\
&&\int d^2k_\perp
\frac{\theta(M^2_{30}-\Lambda^2)}
{M^2_3 -M_{30}^2}\;\Gamma_\Lambda(x,\vec k_\perp).
\label{eqn:fad}
\end{eqnarray}
 
We already studied the two- and the three-body bound
states as a function of the strength $\lambda$ for different values of the cut-off parameter and the stability of the three-body problem has been investigated~\cite{Mattiello:TN03}.

Choosing  $m=336$ MeV for the quark mass we fit our results to the proton mass. This determines the function $\lambda(\Lambda)$ and in this way the model is parameterized by the input $\Lambda$ only. 
In the following we use $\Lambda=4m$ and $\Lambda=8m$ for the calculation in a quark medium.

%

\section{Three-quark correlations in medium}
The Dyson approach allows us to consistently derive relativistic few-body equations for particles embedded in a medium of
both finite temperature and finite density. They systematically include the effects of self energy corrections $m=m(T,\mu)$ and Pauli blocking factors, given in terms of the Fermi distribution function. On the light front it reads
\be
f^+(k^+,\vec k_\perp)=\left(\exp\frac{1}{T}\left[\left(\frac{\vec k^2_\perp+m^2}{2k^+}+\frac{k^+}{2}-\mu\right)\right]+1\right)^{-1}.
\label{eqn:Fermi}
\ee
The dependence of the constituent quark mass on the medium has been calculated within the Nambu-Jona-Lasinio model on the light front~\cite{Mattiello:chiral}.
The two-body propagator is formally given by eq.(\ref{eqn:tau}), where the regularized loop-integral is modified as following
\be
B_\gL(M_2)=-\frac{i}{2(2\pi)^3} \int_{\gL^2\le M_{20}^2} \frac{dx d^2k_\perp}{x(1-x)}
\frac{1-2f^+(x,\vec k^2_\perp)}{M_2^2-M_{20}^2}.
\label{eqn:Bmed}
\ee
The pole of the two-body propagator at finite temperatures and densities determinates the mass of the two-quark system in quark matter.\\
The three-quark equation becomes
\begin{eqnarray}\label{eqn:med3}
\Gamma_\Lambda(y,\vec q_\perp) &= &\frac{i}{(2\pi)^3}\ t_\Lambda(M_2)
\int_0^{1-y} \frac{dx}{x(1-y-x)}\\
&&\hspace*{-1.5cm}\int_{\gL^2\le M_{30}^2} d^2k_\perp
\frac{
1-f^+(x,\vec k_\perp)-f^+(y,(\vec k+\vec q)_\perp )}
{M^2_3 -M_{30}^2}\;\Gamma_\Lambda(x,\vec k_\perp)\nonumber.
\end{eqnarray}

\begin{figure}[b]
\begin{minipage}[t]{0.48\textwidth}
\psfig{figure=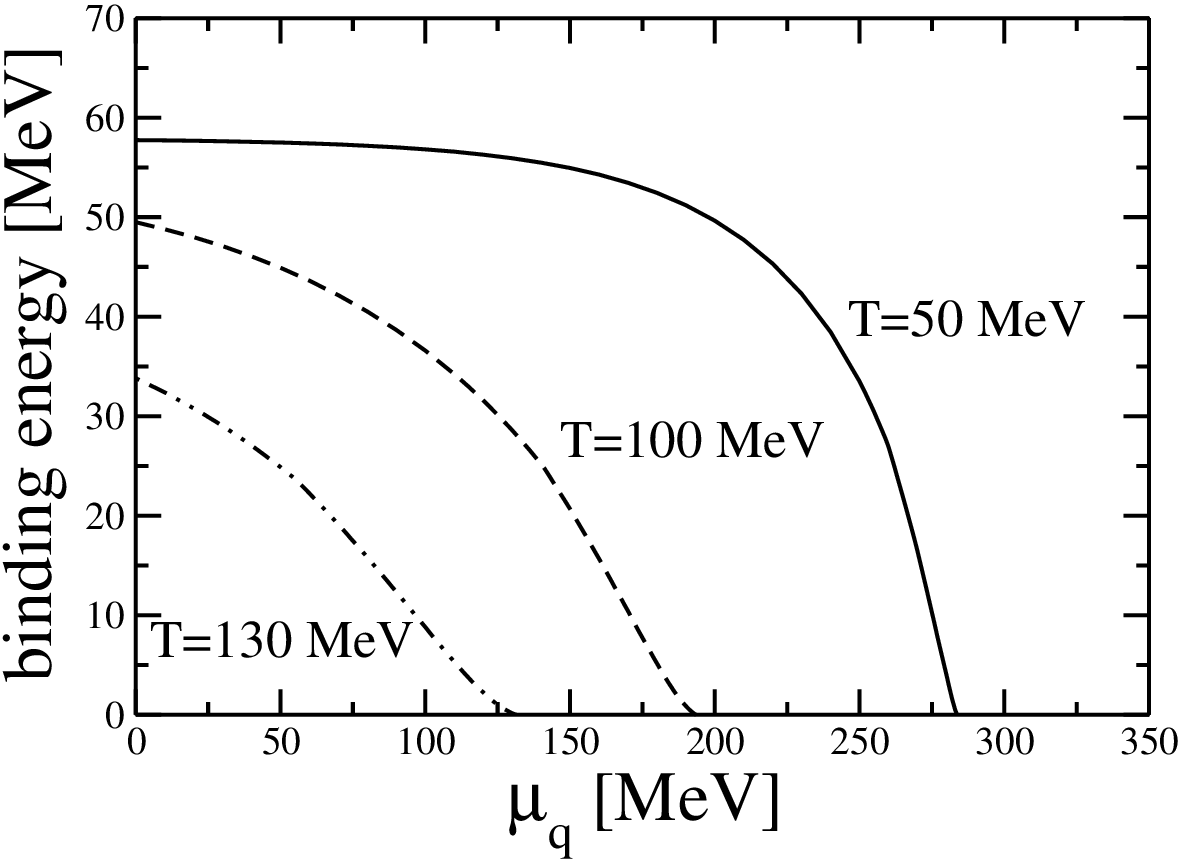,width=0.9\textwidth}
\caption{Binding energy of the three-quark bound state for $\Lambda=4m$ at different temperatures as indicated.\label{fig:BE}}
\end{minipage}
\hfill
\begin{minipage}[t]{0.48\textwidth}
\psfig{figure=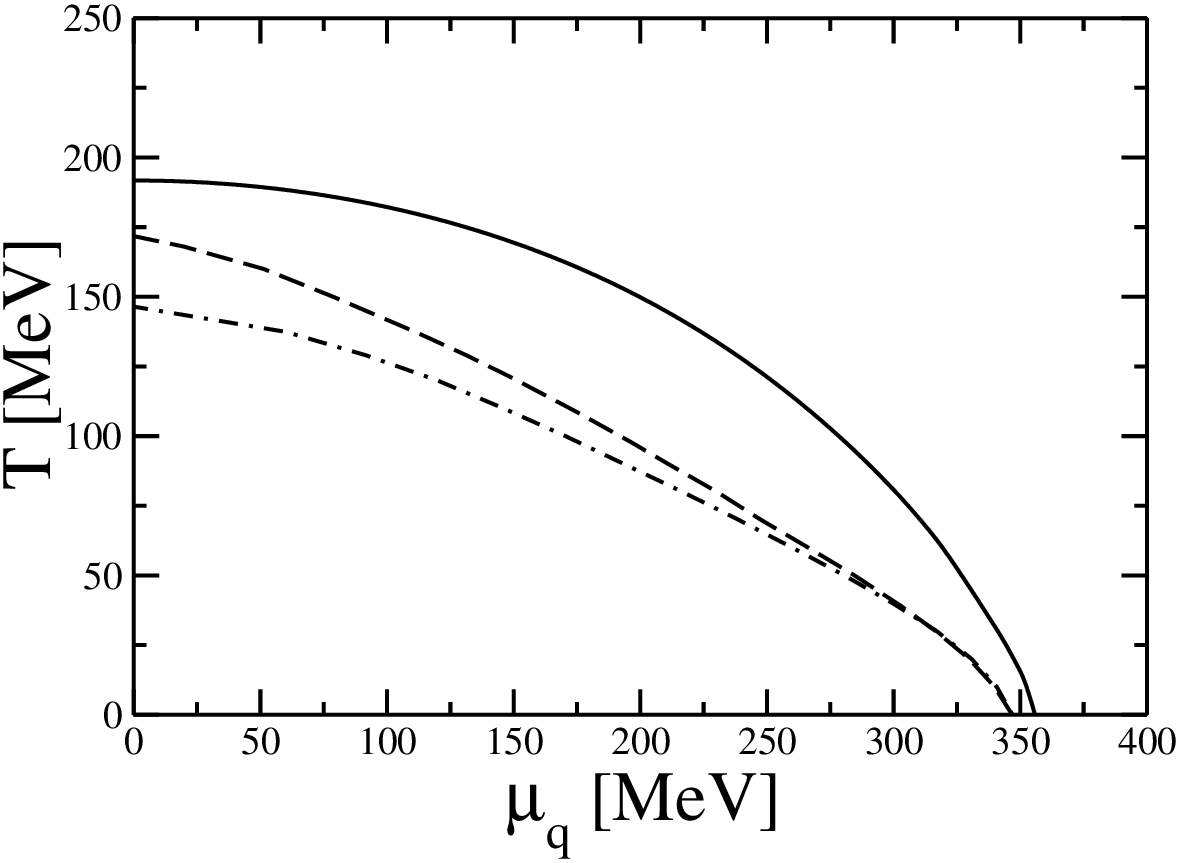,width=0.9\textwidth}
\caption{Dissociation lines of the three-quark bound state as for different invariant cut-offs: $\Lambda=4m$ (dash) and $\Lambda=8m$ (dash-dot). The solid line shows the chiral phase transition.\label{fig:phase}}
\end{minipage}
\end{figure}
The solution of eq.(\ref{eqn:med3}) allows us to calculate the three-quark binding energy at finite temperatures and the chemical potentials for the different cut-offs defined as
\be
B_3(T,\mu)=m(T,\mu)+M_{2B}(T,\mu)-M_{3B}(T,\mu).
\ee
In Fig.~\ref{fig:BE} the binding energy as function of the chemical potential for constant values of the temperature using $\Lambda=4m$ is shown.
$B_3$ becomes smaller by increasing chemical potential for a constant value of the temperature.
We can calculate at which temperature and chemical potential the binding energy goes to zero and therefore the three-quark bound states disappear. The values of $T$ and $\mu$ for which the proton dissociation occurs define the Mott lines and are shown in Fig.~\ref{fig:phase}.
Their behavior qualitatively reflects the chiral phase transition given by the solid line and calculated within Nambu-Jona-Lasinio model on the light front~\cite{Mattiello:chiral}.
The comparison between Mott line and the chiral phase transition shows that the nucleon dissociation occurs in the broken phase.
The difference between these two transition may indicate the confinement-deconfinement region. 
At low temperatures the dependence on the cut-off is mild, but at zero density the different $\Lambda$ give different values for the transition temperature.

\section{Conclusion} 
We have presented relativistic equations of the three-body problem using a zero range interaction and derived consistent relativistic three-quark equations at finite density and temperature.
We have investigated the properties of the three-quark bound states in-medium, i.e the binding energy. In particular we have calculated the nucleon dissociation line and compare it with the chiral phase transition. Further analysis including a treatment of
spins is left for future investigations.
\begin{acknowledge}
We are grateful to S. Strau\ss, T. Frederico and H.J. Weber for the fruitful collaboration.
We thank the organizers of the Light Cone work shop for a very inspiring meeting.
Work supported by Deutsche Forschungsgemeinschaft.
\end{acknowledge}


\begin{thebibliography}{99}
\bibitem{Karsch:2000vy}
F.~Karsch,
Nucl.\ Phys.\ Proc.\ Suppl.\ {\bf 83}, 14 (2000)
\bibitem{Katz:NPPS129}
S.D.~Katz,
Nucl.\ Phys.\ Proc.\ Suppl.\ {\bf 129},60 (2004)
\bibitem{Alford:2001dt}
M.~Alford,
hep-ph/0102047.
\bibitem{Dirac:RMP21}
P.A.M.~Dirac,
Rev. Mod. Phys. {\bf 21}, 392 (1949) 
\bibitem{Mattiello:chiral}
M.~Beyer, S.~Mattiello, T.~Frederico and H.J.~Weber,
hep-ph/0310222 
\bibitem{Beyer:2001bc}
M.~Beyer, S.~Mattiello, T.~Frederico and H.J.~Weber,
Phys.\ Lett.\ B {\bf 521}, 33 (2001)

\bibitem{Frederico:1992uw}
T.~Frederico, Phys.\ Lett.\ B {\bf 282}, 409 (1992)
\bibitem{Carbonell:2002qs}
J.~Carbonell and V.A.~Karmanov,
Phys.\ Rev.\ D {\bf 67}, 052001 (2003)
\bibitem{Mattiello:TN03}
S.~Mattiello, Few Body Syst.\  {\bf 34}, 119 (2004) and refs. therein 




\end{thebibliography}
\end{document}